\documentclass{emulateapj}
\usepackage{apjfonts}
\lefthead{HAN} 
\righthead{NEAR-FIELD MICROLENSING}

\newcommand{\varphivec}{\mbox{\boldmath $\varphi$}}
\newcommand{\thetavec}{\mbox{\boldmath $\theta$}}

\begin{document}
\title{Near-Field Microlensing from Wide-Field Surveys}

\author{Cheongho Han}
\affil{
Program of Brain Korea 21,
Institute for Basic Science Research,
Department of Physics,\\
Chungbuk National University, Chongju 361-763, Korea;
cheongho@astroph.chungbuk.ac.kr}



\begin{abstract}
We estimate the rate of near-field microlensing events expected 
from all-sky surveys and investigate the properties of these events.  
Under the assumption that all lenses are composed of stars, our 
estimation of the event rate ranges from $\Gamma_{\rm tot}\sim 0.2 
\ {\rm yr}^{-1}$ for a survey with a magnitude limit of $V_{\rm lim}
=12$ to $\Gamma_{\rm tot} \sim 20\ {\rm yr}^{-1}$ for a survey with 
$V_{\rm lim}=18$.  We find that the average distances to source stars 
and lenses vary considerably depending on the magnitude limit, while 
the dependencies of the average event time scale and lens-source 
transverse speed are weak and nearly negligible, respectively.  
We also find that the the average lens-source proper motion of 
events expected even from a survey with $V_{\rm lim}=18$ would be 
$\langle \mu\rangle \gtrsim 40\ {\rm mas}\ {\rm yr}^{-1}$, implying 
that the source and lens of a significant fraction of near-field 
events could be resolved from high-resolution follow-up observations.  
From the investigation of the variation of the event characteristics 
depending on the position of the sky, we find that the average 
distances to source stars and lenses become shorter, the lens-source 
transverse speed increases, and the time scale becomes shorter as 
the the galactic latitude of the field increases.  Due to the 
concentration of events near the galactic plane, we find that 
$\gtrsim 50\%$ of events would be detected in the field with 
$b\leq 20^\circ$.
\end{abstract}


\section{Introduction}

The concept of one star gravitationally amplifying the light from 
another background star was first considered by \citet{einstein36},
although he concluded that the chance to observe this phenomenon 
(gravitational microlensing) would be very low.  With the advance 
in technology, however, detections of microlensing events became 
feasible from systematic searches.  The first detections were 
reported in the early 1990s \citep{alcock93, aubourg93, udalski93}.  
To date, lensing events are routinely detected with a rate of 
$\gtrsim 500$ events per season and the total number of detections 
now exceeds 3000.  To maximize the number of detections, these 
searches have been and are being conducted toward very dense 
star fields of nearby galaxies such as the Magellanic Clouds 
\citep{alcock00, afonso03} and M31 \citep{ansari99, calchinovati05, 
uglesich04, dejong04, riffeser03}, and galactic center 
\citep{alcock01, hamadache06, sumi06, bond01}.

Recently, a detection of a lensing event occurred on a nearby 
star was reported by \citet{gaudi07} and \citet{fukui07}.  The 
star (GSC 3656-1328) is located about 1 kpc from the sun in the 
disk of the Milky Way.  This detection along with the advent of 
transient surveys capable of covering a very wide field with 
high cadence such as  ASAS \citep{szczygie07}, ROTSE \citep{akerlof03}, 
TAROT and ARAGO \citep{boer01}, and Pan-Starrs \citep{hodapp04}
have drawn attention of many researchers in microlensing community 
on the feasibility of near-field microlensing surveys.  If near-field 
events could be detected from such surveys, they would be able to 
provide precious information about the matter distribution around 
the sun including dark objects.  In addition, these surveys might 
enable detections of close planets  with the microlensing technique 
that has demonstrated its capability in detecting distant planets 
\citep{bond04, udalski05, beaulieu06, gould06}.

The probability of lensing occurring on nearby stars has been  
estimated by \citet{colley95} and \citet{nemiroff98}.  The 
estimation of \citet{colley95} focused on only events that could 
be observable with naked eyes, i.e.\ with a magnitude limit 
$V_{\rm lim}\sim 6$.  \citet{nemiroff98} estimated the lensing 
probability expected from surveys with various magnitude limits.  
However, he estimated the probability in terms of the average 
number of stars undergoing lensing magnification at a moment, 
not in terms of events per year, by using the combination of the 
star-count information and optical depth to lensing.  The optical 
depth to lensing represents the probability that any given star 
is microlensed with a magnification $A \geq 1.34$ at any given time.  
This optical depth is dependent on the mass distribution but it is 
independent of the mass of the individual lensing objects.  Therefore,
the probability based on the optical depth does not provide detailed 
information about the physical parameters of lens systems such as 
the lens mass, locations of the lens and source, and their relative 
transverse speed, and the observables of events such as the event 
time scale and source brightness.  For this information, additional 
modelling of the mass function and velocity distribution of galactic 
matter is required.

In this paper, we extend the works of \citet{colley95} and 
\citet{nemiroff98} to estimate the rate of lensing events in 
terms of event per year and to obtain the distributions of the 
physical parameters and observables of events expected from 
all-sky lensing surveys with various magnitude limits.  For 
this, we conduct Bayesian simulation of near-field lensing events.

The paper is organized as follows.  In \S\ 2, describe the details 
of the simulation.  In \S\ 3, we present the resulting event rate 
and distributions of the physical parameters of lens systems and 
observables of lensing events.  We also investigate the variation 
of the event characteristics and lensing probability depending on 
the position of the sky.  We analyze the tendencies found in the 
distribution and explain the reasons for these tendencies.  We 
summarize the result and conclude in \S\ 4.

\section{Lensing Simulation}

To investigate the rate and properties of near-field lensing events 
expected to be detected from wide-field surveys, we conduct Bayesian 
simulation of these events.  The basic scheme of the simulation is as 
follows.  
\begin{enumerate}
\item 
We first produce source stars on the sky that can be seen from a 
survey with a given magnitude limit.  We assign the locations on the 
sky and distances to stars based on a mass distribution model of the 
Galaxy.  The stellar brightness is assigned based on a model luminosity 
function of stars considering distances to stars and extinction.
\item
Once source stars are produced, we then produce lenses whose 
locations and masses are assigned based on models of the mass 
distribution and mass function, respectively.  
\item
To estimate the event rate in terms of events per year, it is 
required to model velocity distributions of source stars and lenses.  
The velocity distribution is also required to estimate the distributions 
of the physical parameters of lens systems and the observables of 
lensing events. 
\end{enumerate}
The details 
of the simulation are described in the following subsections.

\subsection{Stellar Distribution}

We model the mass distribution in the solar neighborhood as a 
double-exponential disk of the form
\begin{equation}
\rho(R, z) \propto \exp \left[-\left( {R-R_0\over h_R} + {|z|\over h_z}
\right)\right],
\label{eq1}
\end{equation}
where $R_0$ is the galactocentric distance of the sun and $h_R$ 
and $h_z$ are the radial and vertical scale heights, respectively.  
We adopt $R_0=8$ kpc.  For the radial scale height, we adopt a 
fixed value of $h_R=4.0$ kpc.  For the vertical scale heights, 
on the other hand, we adopt varying values depending on the stellar 
brightness such that $h_z=117$ pc for $M_V\leq 2.0$, 195 pc for
$M_V\leq 3.0$, 260 pc for $M_V\leq 4.0$, 325 pc for $M_V\leq 5.0$,
390 pc for $M_V\leq 6.0$, 455 pc for $M_V\leq 7.0$, and 520 pc for 
$M_V >   7.0$.

We model the luminosity function of stars by adopting that of 
stars in the solar neighborhood presented in \citet{binney98}.  
This luminosity function is constructed by combining the data 
published in \citet{allen73} for $M_V\leq 0$, in \citet{jahreiss83} 
and \citet{kroupa90} for $M_V > 0$.

Extinction is modeled such that the stellar flux decreases 
exponentially with the increase of the dust column density, 
i.e.\ 
\begin{equation}
A_V=-2.5 \log [\exp(-k \Sigma_{\rm d})],
\label{eq2}
\end{equation}
where $\Sigma_{\rm d}$ is the dust column density and $k$ is a 
proportional constant.  The dust column density $\rho_{\rm d}$ is 
computed on the basis of an expontial dust distribution model, i.e.
\begin{equation}
\Sigma_{\rm d} = \int_0^{D_{\rm S}} 
\rho_{\rm d}(\ell) d\ell,\qquad 
\rho_{\rm d}\propto \exp\left(-{|z|\over h_{{\rm d},z}}\right),
\label{eq3}
\end{equation}
where the integral is along the line of sight toward the source 
star and $h_{{\rm d},z}$ is the vertical scale height of the dust 
distribution.  We adopt $h_{{\rm d},z}=150$ pc.

With these models, we produce source stars with their distances 
from the sun and locations on the sky assigned based on the mass 
distribution model and their absolute magnitudes allocated based 
on the luminosity function.  Then, the apparent magnitude is computed 
based on the distance to the star and the amount of extinction.  
We set the normalization of the stellar number density and extinction 
so that the star count result from our simulation matches the star 
count data as a function of galactic latitude presented in 
\citet{allen98}.

\subsection{Lens Distribution and Mass Function}

In our simulation, we assume that only stars are responsible for 
lensing events and no MACHOs or stellar remnants are taken into 
consideration.  For the mass function of lens matter, we adopt
the model of \citet{gould00} that is constructed based on the 
observations of \citet{zoccali00}.  The model has a double power-law 
distribution of the form 
\begin{equation}
{dN\over dm} = k \left( {m\over m_c}\right)^\gamma;\qquad
\gamma =
\cases{
-2.0  & for $m>m_{\rm c}$, \cr
-1.3  & for $m<m_{\rm c}$,\cr
} 
\label{eq4}
\end{equation}
where the turning-point mass is $m_{\rm c}=0.7\ M_\odot$.  The 
distance to the individual lenses are assigned from the same mass 
distribution model as that of source stars in equation~(\ref{eq1}).  
Since the lens is a star, its flux contributes to the observed flux.  
In this case, the flux from the lens works as blended flux to the
observed flux.  Then, the apparent magnification of an event 
appears to be lower than actual values \citep{nemiroff97}.  However, 
considering that source stars are bright enough to be monitored 
with small telescopes while most lenses are faint stars, the lens 
contribution to the observed flux is in most cases negligible
\citep{han98}.

\begin{figure}[t]
\epsscale{1.1}
\plotone{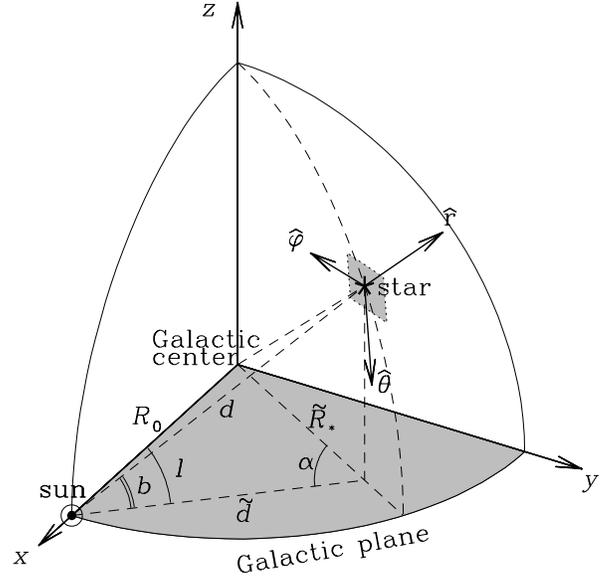}
\caption{\label{fig:one}
Position of a star as seen from the sun.  $(l,b)$ are the galactic 
longitude and latitude of the star, $d$ is the distance to the star, 
and $R_0$ is the galactocentric distance of the sun.  The vectors
$\hat{\bf r}$, $\hat{\varphivec}$, and $\hat{\thetavec}$
represent the unit vectors in the spherical coordinates centered 
at the position of the sun.  The projected velocity of the star 
corresponds to ${\bf v}_\perp = v_\varphi \hat{\varphivec} +
v_\theta  \hat{\thetavec}$.
}\end{figure}

\subsection{Velocity Distribution}

We model the velocity distribution as gaussian, i.e.\ 
\begin{equation}
f({\bf v}) \propto \exp \left[ {({\bf v}-\bar{\bf v})^2\over 
2\sigma^2}\right].
\label{eq5}
\end{equation}
The adopted means and standard deviations of the distributions of 
the individual velocity components in the galactocentric cylinderical 
coordinates are $(\bar{v}_R,\bar{v}_\phi,\bar{v}_z) =(0,220,0)\ 
{\rm km}\ {\rm s}^{-1}$ and $(\sigma_R,\sigma_\phi, \sigma_z)=
(38,25,20)\ {\rm km}\ {\rm s}^{-1}$, respectively, i.e.\ a 
flat-rotating disk with gaussian velocity dispersion.  For lensing, 
only the projected velocity, ${\bf v}_\perp$, as seen from the 
observer are related to lensing events.  For lensing simulation, 
then, it is required to convert the velocity components into those 
in the spherical coordinates that are centered at the position of 
the sun.  This conversion is done by the relation
\begin{equation}
\left( {\begin{array}{c}
v_r \\ v_\varphi \\ v_\theta 
\end{array}} \right)  
= 
\left({\begin{array}{rrr}
 \cos b & 0 & \sin b \\
 0      & 1 & 0 \\
-\sin b & 0 & \cos b 
\end{array}}\right)
\left({\begin{array}{rrr}
 \cos\alpha & \sin\alpha  & 0 \\
-\sin\alpha & \cos\alpha  & 0 \\
 0          & 0           & 1 
\end{array}}\right)
\left( {\begin{array}{c}
v_R \\ v_\phi \\ v_z
\end{array}} \right),
\label{eq6}
\end{equation}
where $b$ is the galactic latitude of the star and $\alpha$ is 
the angle between the two lines connecting the sun, the projected 
position of the star on the galactic plane, and the galactic center 
(see Figure~\ref{fig:one}).  The angle $\alpha$ is related to the 
galactic longitude of the star ($l$), galactocentric distance of 
the sun, and the projected distance to the star on the galactic 
plane from the sun ($\tilde{d}$) by
\begin{equation}
\alpha=\sin^{-1} \left({R_0\sin l \over \tilde{R}_\star}\right);
\ \ \ \ 
\tilde{R}_\star = (R_0^2+\tilde{d}^2-2R_0\tilde{d}\cos l)^{1/2},
\label{eq7}
\end{equation}
where $\tilde{R}_\star$ is the projected distance to the star from 
the galactic center (see Figure~\ref{fig:one}).  Then, the projected 
velocity of a star depends not only on the galactic coordinates 
$(l,b)$ but also on the distance to the star $d=\tilde{d} \sec b$, 
i.e.\ ${\bf v}_\perp = (v_\varphi, v_\theta) = {\bf v}_\perp (l,b,d)$.  
With the projected velocities of the lens, ${\bf v}_{\rm L}=
{\bf v}_\perp(D_{\rm L})$, and source, ${\bf v}_{\rm L} = 
{\bf v}_\perp (D_{\rm S})$,  the relative lens-source transverse 
velocity is computed by
\begin{equation}
{\bf v}_{\rm t} = {\bf v}_{\rm L} - 
\left[ {\bf v}_{\rm S}
\left( {D_{\rm L}\over D_{\rm S}}\right) 
+ {\bf v}_{\rm O} \left( {D_{\rm S}-D_{\rm L}\over D_{\rm S}}
\right) \right],
\label{eq8}
\end{equation}
where ${\bf v}_{\rm O}$ represents the projected velocity of the 
observer and $D_{\rm L}$ and $D_{\rm S}$ are the distances to the 
lens and source, respectively.  We note that the observer, lens, 
and source are all in rotation with the same rotation speed, and 
thus the contribution of the rotation to the transverse speed is 
negligible.  Then, the lens-source transverse motion is mostly 
caused by the dispersion of the rotation-subtracted residual velocity.

\subsection{Event Rate}

Once the source, lens, and their relative speed are chosen, its 
contribution to the event rate is computed by
\begin{equation}
\Gamma \propto n(l,b,D_{\rm S}) n(l,b,D_{\rm L})D_{\rm S}^2 
\sigma v_{\rm t},
\label{eq9}
\end{equation}
where $n(l,b,d)$ represents the number density of stars at the 
position $(l,b,d)$ and $\sigma$ is the cross-section of the 
lens-source encounter.  Here the factor $D_{\rm S}^2$ is included 
to account for the increase of the number of source stars with the 
increase of the distance.  Under the definition of a lensing event 
as the approach of the source star within the Einstein ring of the
lens (equivalently, source flux is magnified with a magnification 
$A\geq 1.34$), the lensing cross-section is set as $\sigma=2 r_{\rm E}$, 
where $r_{\rm E}$ is the physical Einstein ring radius.  The Einstein 
radius is related to the physical parameters of the lens system by
\begin{equation}
r_{\rm E}=\left( {4GM\over c^2}\right)^{1/2}
\left[ {D_{\rm L}(D_{\rm S}-D_{\rm L})\over D_{\rm S}}\right]^{1/2},
\label{eq10}
\end{equation}
where $M$ is the mass of the lens.  We assume that the survey is 
conducted all through the year\footnote{This requires a network of 
wide-field telescopes distributed over the Earth.} and events are 
detected with 100\% efficiency.

\begin{deluxetable}{crrc}
\tablecaption{Rates of Microlensing Events\label{table:one}}
\tablewidth{0pt}
\tablehead{
\colhead{$V_{\rm lim}$} &
\colhead{$N_{\star,{\rm tot}}$} &
\colhead{$\langle \tau\rangle$} &
\colhead{$\Gamma_{\rm tot}$ (${\rm yr}^{-1}$)} 
}
\startdata
12  & $0.14\times 10^7$  & $0.16\times 10^{-8}$  &  0.18\\
14  & $0.65\times 10^7$  & $0.32\times 10^{-8}$  &  0.61\\
16  & $2.4\times 10^7$   & $0.60\times 10^{-8}$  &  3.86\\ \smallskip
18  & $8.0\times 10^7$   & $1.0\times 10^{-8}$   & 21.20 
\enddata
\tablecomments{ 
Rates of near-field microlensing events, $\Gamma_{\rm tot}$, 
expected from all-sky surveys with various magnitude limits, 
$V_{\rm lim}$.  Also listed are the total number of stars that 
can be monitored from the surveys, $N_{\star,{\rm tot}}$, and the 
average optical depth to lensing, $\langle \tau \rangle$.}
\end{deluxetable}

\begin{figure}[ht]
\epsscale{1.15}
\plotone{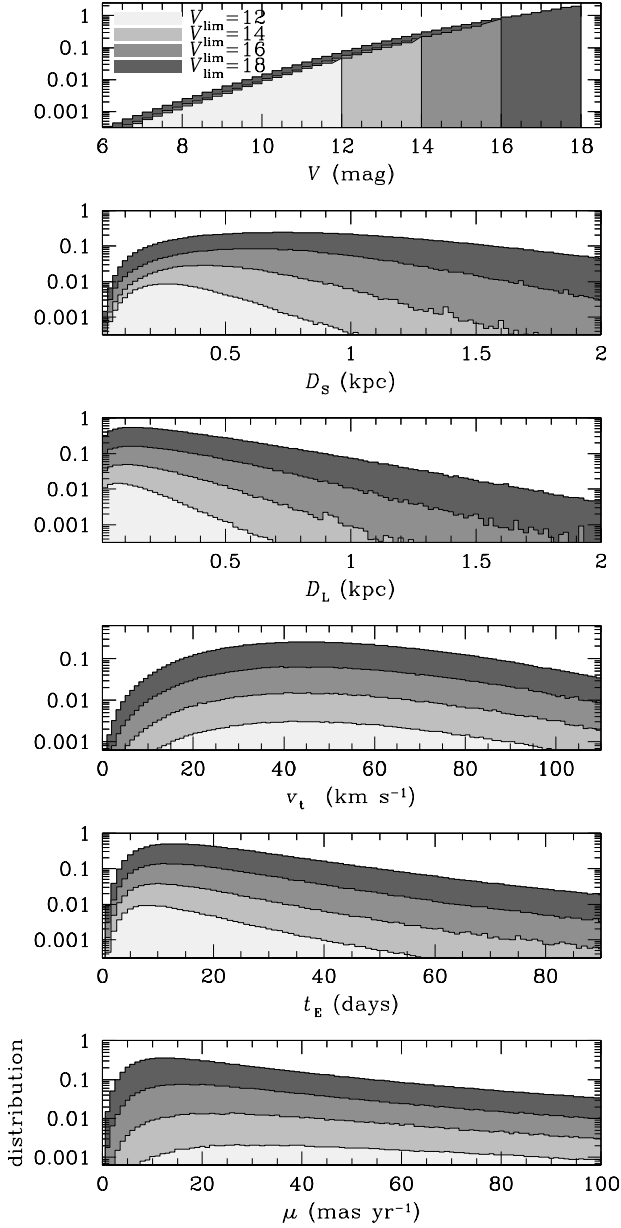}
\caption{\label{fig:two}
Distributions of lens parameters and observables of near-filed 
microlensing events expected from all-sky surveys with various 
magnitude limits, $V_{\rm lim}$.  From the top, the individual panels 
represent the distributions of the source star brightness ($V$), 
distances to the source star ($D_{\rm S}$) and lens ($D_{\rm L}$), 
transverse speeds ($v_{\rm t}$), event time scale ($t_{\rm E}$), and 
lens-source relative proper motions ($\mu$), respectively.  The 
mean values of the parameters and observables are listed in 
Table~\ref{table:two}.
}\end{figure}

\section{Results}

In Table~\ref{table:one}, we list the total event rate, 
$\Gamma_{\rm tot}$, expected from surveys with various magnitude 
limits.  Also listed in the table are the total number of stars 
that can be monitored from the surveys, $N_{\star,{\rm tot}}$, 
and the average optical depth to lensing, $\langle \tau \rangle$.  
The optical depth is determined from the mass distribution model by
\begin{equation}
\tau(l,b)={\int_0^\infty dD_{\rm S} n(l,b,D_{\rm S}; V\leq V_{\rm lim})
D_{\rm S}^2 \int_0^{D_{\rm S}} dD_{\rm L} n(l,b,D_{\rm L}) r_{\rm E}^2 
\over 
\int_0^\infty dD_{\rm S} n(l,b,D_{\rm S}; V\leq V_{\rm lim})D_{\rm S}^2},
\label{eq11}
\end{equation}
where the factor $\int_0^\infty dD_{\rm S} n(l,b,D_{\rm S};
V\leq V_{\rm lim})D_{\rm S}^2$ 
corresponds to the number of stars brighter than the magnitude 
limit toward the observational field, $N_\star$, and the other 
factor $\int_0^{D_{\rm S}} dD_{\rm L} n(l,b,D_{\rm L}) r_{\rm E}^2$ 
is the accumulation of the area occupied by the Einstein rings of 
the individual lenses along the line of sight toward the source star.  
We note that although there exists limitation in the brightness of 
source stars ($V< V_{\rm lim}$), there is no restriction in the lens 
brightness.  The presented value in Table~\ref{table:one} are the 
mean optical depth averaged over the whole sky, i.e.\ $\langle \tau 
\rangle = (4\pi)^{-1} \int_{-180^\circ}^{180^\circ}dl 
\int_{-90^\circ}^{90^\circ}\tau(l,b) \sin b\ db$, and the total 
number of stars over the entire sky, i.e.  $N_{\star,{\rm tot}} = 
\int_{-180^\circ}^{180^\circ}dl \int_{-90^\circ}^{90^\circ} 
N_\star(l,b) \sin b\ db$.  We find that our estimation of the 
optical depth matches well with the estimation of \citet{nemiroff98}.  
The multiplication $N_{\star,{\rm tot}} \times \langle \tau\rangle$ 
represents the number of stars undergoing lensing magnification 
at a given moment, which was presented in \citet{nemiroff98}.  We 
also find a good match between his and our estimations.  The estimated 
event rate ranges from $\Gamma_{\rm tot}\sim 0.2$ per year for a 
survey with a magnitude limit of $V_{\rm lim}=12$ to $\Gamma_{\rm tot}
\sim 20$ per year for a survey with $V_{\rm lim}=18$, confirming the 
result of \citet{nemiroff98} that the lensing probability rapidly 
increases with the increase of the magnitude limits.  Two factors 
contribute to the increase of the event rate with the increase of 
the magnitude limit.  The first factor is the increase of the number 
of source stars and the other factor is the increase of the line 
of sight toward source stars and thus increase of the optical depth.  
We find that the former factor is more important.

In Figure~\ref{fig:two}, we present the distributions of lens 
parameters and the observables of events.  From the top, the individual 
panels represent the distributions of the source star brightness, 
distances to source stars and lenses, lens-source transverse 
speed, event time scale, and lens-source relative proper motion, 
respectively.  The event time scale is determined by the Einstein 
time scale, which is required for the source to transit the Einstein 
radius of the lens, i.e.\ $t_{\rm E}=r_{\rm E}/v$.  The proper motion 
corresponds to $\mu=\theta_{\rm E}/t_{\rm E}$, where $\theta_{\rm E}
=r_{\rm E}/D_{\rm L}$ is the angular Einstein radius.  In 
Table~\ref{table:two}, we also present the average values of the 
lens parameters.  From Figure~\ref{fig:two} and Table~\ref{table:one} 
and \ref{table:two} we find the following tendencies.
\begin{enumerate}
\item 
The distributions of $D_{\rm S}$ and $D_{\rm L}$ vary considerably 
depending on the magnitude limit.  The trend, as expected, is that 
the distances to the source and lens become larger as fainter stars 
are monitored.
\item  
On the other hand, the dependency of the distribution of $t_{\rm E}$ 
on the magnitude limit is not very strong and the dependency of the 
velocity distribution is nearly negligible.  
\item 
With the increase of the magnitude limit, the distance to the lens 
increases while the transverse speed remains nearly the same.  As a 
result, the average proper motion $\mu=\theta_{\rm E} /t_{\rm E}
=v_{\rm t}/D_{\rm L}$ of events decreases as the magnitude limit 
increases.  However, we note that even for surveys with $V_{\rm lim}
\sim 18$, the mean proper motion of events is $\langle \mu \rangle 
\gtrsim 40\ {\rm mas}\ {\rm yr}^{-1}$, which is much larger than the 
typical value of $5\ {\rm mas}\ {\rm yr}^{-1}$ of galactic bulge 
events.  Then, the lens and source of a significant fraction of 
near-field events can be resolved from follow-up observations by 
using high-resolution instrument such as the {\it Hubble Space 
Telescope} conducted several years after the peak of the magnification.  
This not only enables the measurement of the proper motion but also 
helps to identify the lens.
\end{enumerate}

\begin{deluxetable}{cccccc}
\tablecaption{Average Values of Lens Parameters\label{table:two}}
\tablewidth{0pt}
\tablehead{
\colhead{$V_{\rm lim}$} &
\colhead{$\langle D_{\rm S} \rangle$} &
\colhead{$\langle D_{\rm L} \rangle$} &
\colhead{$\langle v_{\rm t}\rangle$} &
\colhead{$\langle t_{\rm E} \rangle$} &
\colhead{$\langle \mu \rangle$} \\
\colhead{(mag)} &
\colhead{(kpc)} &
\colhead{(kpc)} &
\colhead{$({\rm km}\ {\rm s}^{-1})$} &
\colhead{(days)} &
\colhead{(${\rm mas}\ {\rm yr}^{-1}$)} 
}
\startdata
 12  &  0.39 &  0.19 & 55.3  &  18.7 &  71.5   \\
 14  &  0.56 &  0.27 & 55.4  &  21.9 &  58.5   \\
 16  &  0.78 &  0.36 & 55.6  &  25.0 &  49.1   \\  \smallskip
 18  &  1.03 &  0.44 & 56.0  &  27.4 &  44.1   
\enddata
\tablecomments{ 
Average values of the lens parameters and observables of near-field 
microlensing events expected from wide-field surveys with various 
magnitude limits, $V_{\rm lim}$.} 
\end{deluxetable}

\begin{figure*}[t]
\epsscale{0.8}
\caption{\label{fig:three}
Variation of the characteristics and probability of near-field 
microlensing events depending on the position of the sky.  From 
the top, the panels in each column represent the distributions of 
the average distances to source star ($D_{\rm S}$)  and lens 
($D_{\rm L}$), lens-source transverse speed ($v_{\rm t}$), event 
time scale ($t_{\rm E}$), average number density of stars in the 
field ($N_\star$), optical depth ($\tau$), and the event rate per 
unit angular area ($\Gamma$).  The panels in each row represent 
the distributions for different magnitude limits. The map is 
centered centered at the galactic center, i.e. $(l,b)=(0^\circ,
0^\circ)$.
}\end{figure*}

Figure~\ref{fig:three} shows the variation of the event characteristics 
and probability of lensing depending on the position of the sky.
From the top, the panels in each column represent the distributions 
of the average distances to source stars and lenses, lens-source 
transverse speed, event time scale, number density of stars in the 
field, optical depth, and the event rate per unit angular area of 
the sky.  The panels in each row represent the distributions for 
different magnitude limits.  The tendencies found from the maps are 
as follows.
\begin{enumerate}
\item
For a survey with a given magnitude limit, the average distances to 
source stars and lenses become smaller as the latitude of the field 
increases.  This is because the number density of stars along the line 
of sight decreases more rapidly as the field is located further away 
from the disk plane.
\item 
The transverse speed increases with the increase of the latitude.
This tendency is due to the elongation of the velocity ellipsoid of 
the rotation-subtracted residual stellar motion.  The major axes of 
the velocity ellipsoid are $(\sigma_R,\sigma_\phi,\sigma_z)=(38,25,20)
\ {\rm km}\ {\rm s}^{-1}$ and thus the ellipsoid is elongated more 
along the galactic plane than along the pole direction.  Then, the 
dispersion of the projected velocity distribution for stars located 
toward the direction of the galactic plane is $\sigma_\perp(b=0^\circ)
=(\sigma_\phi^2+\sigma_z^2)^{1/2} \sim 32 \ {\rm km}\ {\rm s}^{-1}$, 
while the velocity dispersion for stars located toward the galactic 
pole direction is $\sigma_\perp(b=90^\circ) = (\sigma_R^2 + 
\sigma_\phi^2)^{1/2} \sim 45\ {\rm km}\ {\rm s}^{-1}$.  This also 
explains the tendency of the longer time scales for events occurred 
on stars located at a lower latitude.
\item 
With the increase of the stellar number density combined with the 
increase of the optical depth, the event rate increases rapidly 
toward the galactic plane.  We find that $\gtrsim 50\%$ of events 
are expected to be found in the fieled with $b\leq 20^\circ$.
\end{enumerate}

\section{Conclusion}

Instigated by the recent discovery of an event occurred on a nearby 
star, we estimated the rate of near-field lensing events expected 
from all-sky surveys with various magnitude limits.  Under the 
assumption that all lenses are composed of stars, our estimation 
of the event rate ranges from $\Gamma_{\rm tot}\sim 0.2$ per year 
for a survey with a magnitude limit of $V_{\rm lim}=12$ to 
$\Gamma_{\rm tot}\sim 20$ per year for a survey with $V_{\rm lim}=18$, 
confirming the previous result  that lensing probability rapidly 
increases with the increase of the magnitude limit.  The increase 
of the rate is due to two factors which are the extension of the line 
of the sight toward source stars and the increase of the number of 
source stars.  We found that the latter factor is more important.
We also investigated the distributions of the physical parameters 
of lens systems and the observables of these events.  From this, we 
found that although the average distances to source stars and lenses 
vary considerably depending on the magnitude limit, the dependencies 
of the average event time scale and lens-source transverse speed are 
weak and nearly negligible, respectively.  We also found that the 
the average lens-source proper motion of events expected from a 
survey with $V_{\rm lim}=18$ would be $\langle \mu\rangle \gtrsim 
40\ {\rm mas}\ {\rm yr}^{-1}$ and the value further increases as 
the magnitude limit becomes lower.  This implies that the source 
and lens of a typical near-field lensing event can be resolved from 
high-resolution follow-up observations conducted several years after 
the peak of the lensing magnification.  From the investigation of 
the variation of the event characteristics depending on the position 
of the sky, we found that the average distances to source stars and 
lenses become shorter, the lens-source transverse speed increases, 
and the time scale becomes shorter as the the galactic latitude of 
the field increases.  Due to the concentration of events near the 
galactic plane, we found that $\gtrsim 50\%$ of events would be 
detected in the field with $b\leq 20^\circ$.

\acknowledgments 
This work was supported by the research grant of Chungbuk National 
University in 2007.

\end{document}